# On the quantitative optical properties of Au nanoparticles embedded in biological tissue phantoms

J. C. R. Araújo, A. F. G. Monte, R. L. Serrano, W. Iwamoto, A. Antunes, O. Brener, M. Foschini

Institute of Physics, Federal University of Uberlândia, Uberlândia-MG, 38400-902, Brazil

## Abstract

We systematically investigated and quantified how gold (Au) metal nanoparticles (NPs) optical spectra change upon introduction into biological tissue phantoms environment, in which the AuNPs can agglomerate. Quantitative knowledge of how the AuNPs spectra and plasmon resonance wavelength change inside a phantom environment can provide many in-vitro and in-vivo plasmonic NPs-mediated applications. Because the plasmonic properties of metal NPs are dependent on their size, morphology, concentration and local environment, tuning the incident photon wavelength may increase the AuNPs plasmonic properties, on which applications such as plasmonic photothermal therapy and photonic gene circuits are based. Quantitatively analyzing optical absorption and scattering data, we were able to observe changes in the resonance peak positions and breadths, which are related to the distribution of the spherical AuNPs within the chitosan phantoms as a function of the particles size and concentration. The scattered irradiance with the wavelength for embedded AuNPs was registered for the optical scattering mechanisms, decreasing as $\lambda^{-4}$ for Rayleigh scattering, and much more slowly for Mie scattering, $\lambda^{-b}$, with b in the range of 1 to 3.

## Introduction

Metallic nanoparticles (NPs) have become increasingly important in biomedical sciences and photothermal therapy for different reasons. They are able to be excited at their plasmon resonance maximum to cause localized absorption, heating, and eventually destruction of dangerous cells. For AuNPs, specifically, the plasmonic shift is an important characteristic due to its dependence on the interaction of the metal NPs with the environment [1] [2]. Plasmonic nanostructures are also employed in *in vitro* studies to observe biological processes through Plasmon Resonance Energy Transfer (PRET) [3] [4] [5]. The resonance of the NPs probe must match the absorption of the molecule of interest for PRET to work. Therefore, understanding the conditions under which it may shift upon cellular uptake would help ensure accuracy of PRET. Plasmonic NPs can be also explored as multicolor labels for molecular imaging [6-10]. However, the magnitude of spectral shift resulting from clustering is far from being fully quantified. For instance, it is still unknown how much of the mismatch with the laser emission wavelength can arise from that shift, and also it is worthy of being considered. In principle, by using the NPs agglomeration effect – strictly associated with the PRET - spherical AuNPs excited to the resonance with visible wavelengths can be also excited for photothermal therapy with infrared wavelengths within the biological transparency window.

Strictly thinking on simulating the effects on a biological tissue, we discuss the quantitative optical properties of AuNPs obtained in both colloidal solution and embedded in organic medium. The main optical parameters such as the optical absorption, scattering coefficients, and the anisotropy factor have an important role for biological tissues and phantoms characterization. We also aimed to study the effects of the particle-size and environment in the plasmonic shift [11]. The chemical method used for growing the NPs suggested a potential solution for developing ways to fabricate NPs with narrower resonances. Also, it was important to consider the possible spectral changes in a phantom environment when designing the plasmon resonance characteristics of NPs antennas. Absorption and scattering coefficients for a set of six samples were presented and discussed. Then, the optical coefficients were extracted in order to compare their changes in colloidal and embed in the phantom. Our results suggested that the AuNPs embedded in chitosan phantoms were suitable for being used as reference standard materials for medical and biophysical applications.

**Samples**

Six AuNPs samples with different particle-size distribution were synthesized by the method based on the standard procedure described by Turkevich [12] and Frens [13]. A standard volume of 50.0 ml of Tetrachloroaurate(III) hydrate, $3.0 \times 10^{-4}$ M, ($HAuCl_4 \cdot xH_2O$, 99.995 %, Sigma-Aldrich) in a solution of deionized water is heated up to the boiling point, and a solution of 0.21 ml of 1.0 wt % Trisodium Citrate ($Na_3C_6H_5O_7$, 99 %, Sigma-Aldrich) is then added as a growing agent. The solution was further boiled for 15 minutes under constant stirring followed by cooling to room temperature for 24 hours to allow for stabilization. This process was then repeated for different volumes of the growing solution, such as: 0.30, 0.50, 0.75, 1.00, and 1.75 ml to attain different particle-sizes distributions chosen in such a way to get high photostability and optical absorption coefficients up to 1.0 $mm^{-1}$.

Phantoms were prepared using an initial solution of gelatin from porcine skin (Type-A, Sigma-Aldrich) prepared at 10 % in weight in distilled water at 40 $^{o}$C. Then, 5.0 ml from this solution was cooled to room temperature together with 2.0 ml from aqueous solution of chitosan (2.0 wt %), prepared with 2.0 % stock solution of acetic acid ($CH_3CO_2H$, 99.7 %, VETEC) in deionized water. The mixture was stirred until homogenization, then 5.0 ml from the colloidal solution containing the AuNPs was added to the mixture. The final mixture was stirred for 3 min, poured into a Petri dish, and left to dry at room temperature for 48 hours.

**Experimental Techniques details**

The AuNPs morphology and size distribution were characterized by Transmission Electron Microscope (TEM), using a Hitachi HT7700 operated at 100 kV acceleration voltage. The absorbance spectra as a function of the AuNP sizes were measured in the colloidal solution as well as in the chitosan phantoms using an UV-VIS spectrometer. Nonetheless, the introduction of the NPs into the chitosan phantoms needed more attention. When dealing with optical measurements in cloudy media, which is the case of phantoms and biological tissues, one of the most practical methodologies to measure and determine their optical properties is by using the inverse-adding doubling method (IAD) [14], and the integrating sphere detection [15,16].

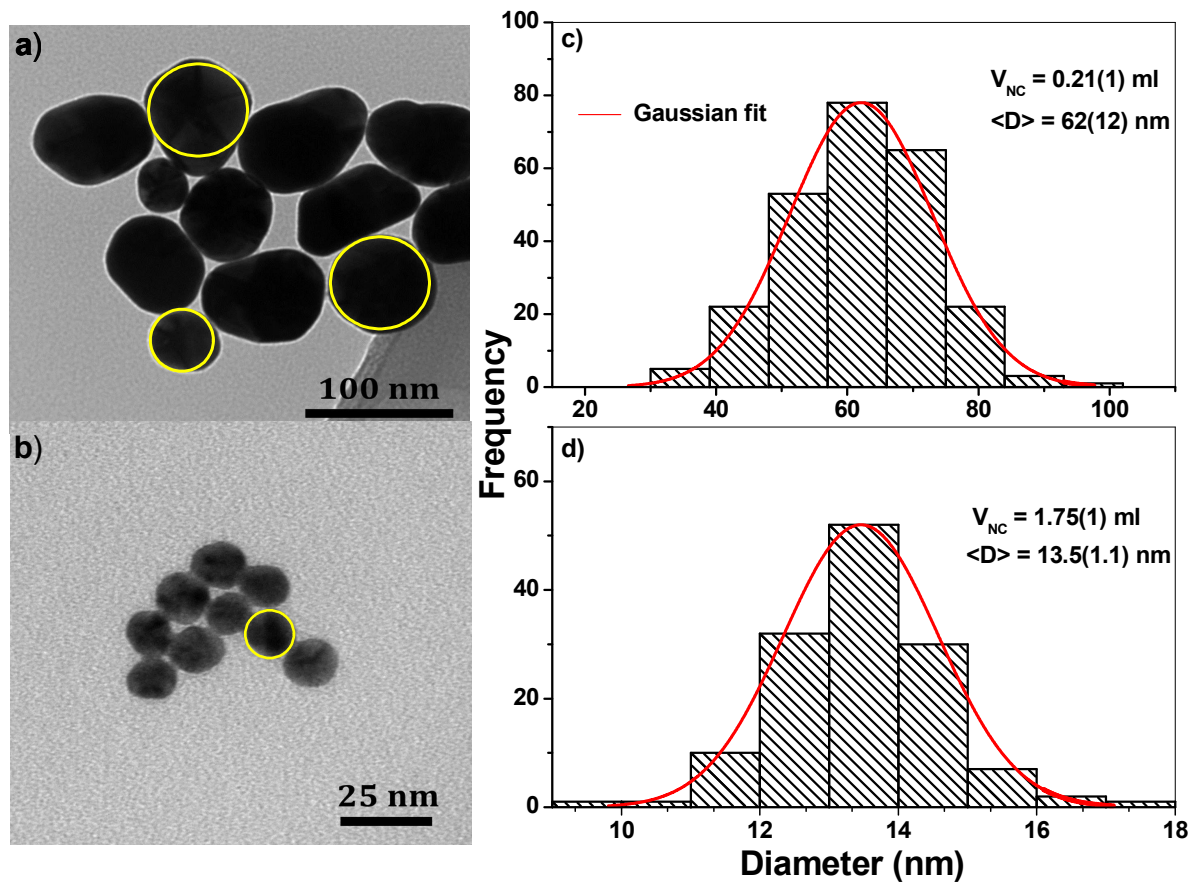

**Fig. 1** Representative TEM images and their histograms of two samples considering the smallest and largest sizes. The particles shape was assumed approximately spherical (yellow circle). Micrographies in (a) and (b) showed the morphology of the AuNPs obtained with different trisodium citrate concentrations ($V_{NC}$). The histograms (c) and (d) revealed that most NPs have an average size of 62(12) nm for $V_{NC}$ = 0.21(1) ml, and 13.5(1.1) nm for $V_{NC}$ = 1.75(1) ml, suggesting that the particle-size variation was related to the concentration of reductant.

**Results**

The nanoparticle-size distribution <D> and the NP morphology were analyzed by TEM images as shown in Fig. 1. The analysis was conducted by fitting the particle-size distribution histogram through Gaussian functions, which considered the particles synthesized with different trisodium citrate concentrations ($V_{NC}$).

In Figure 2, the study of the absorbance spectra for AuNPs in water and embedded in chitosan phantoms were measured at room temperature by scanning the incident wavelength in the range 450-1000 nm. To make clear for the eyes, the absorbance spectra were plotted only for two different sizes, for <D> = 13.5(1.1) nm in Fig. 2(a), and for <D> = 62(12) nm in Fig. 2(b). The absorbance peak maximum ($\lambda_{max}$) from the AuNPs spectra at different NP sizes were also plotted as a function of the average NP size <D>. Their relative absorption cross sections were further examined at different sizes. The beam-media interaction appears to be more effective if one considers the different refractive indexes in water ($n_1$ = 1.00), and chitosan ($n_2$). The spectra showed consistent peaks within the range 550-600 nm. In Fig. 2(c), the decreasing of the volume of trisodium citrate in the mixture ($V_{NC}$) was correlated with the increased of sizes <D>. Not surprisingly, the absorbance peak followed the increase of the AuNP sizes, as in Fig. 2(d).

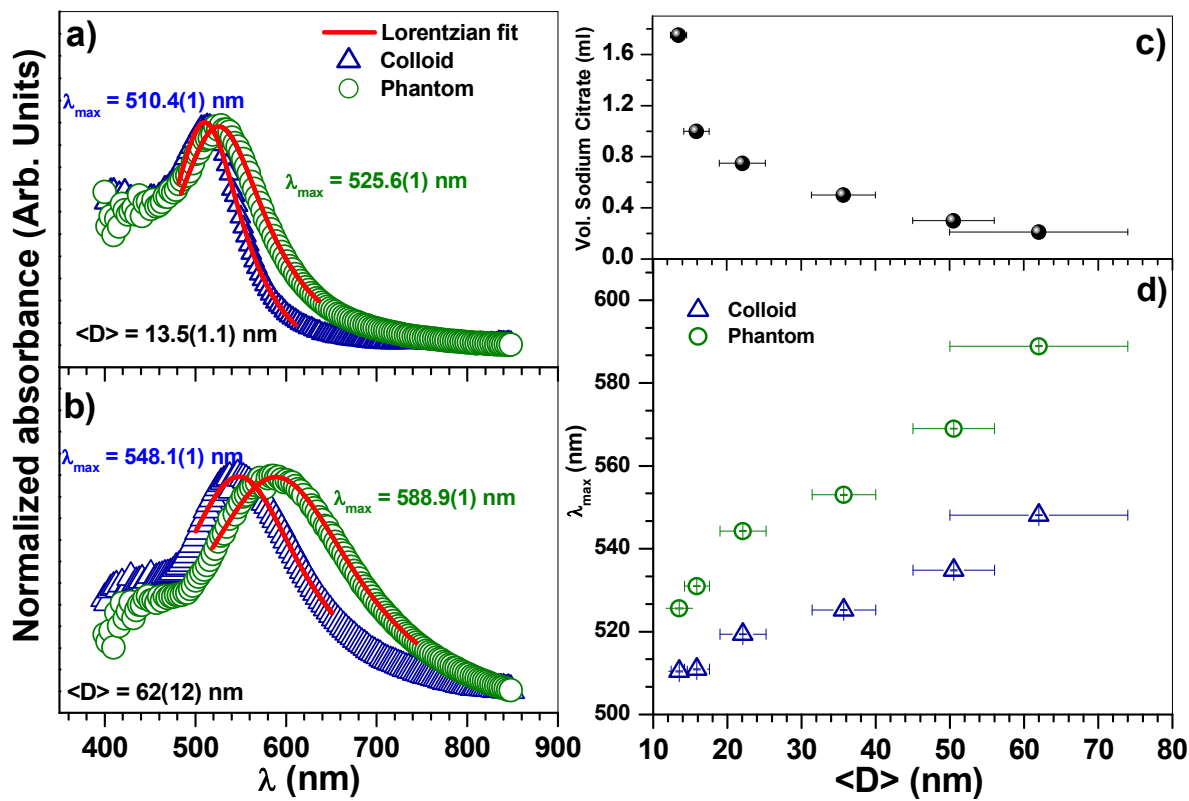

**Fig. 2** (a), (b) Absorbance spectra from AuNPs embedded in water and chitosan phantoms for two different nanoparticle sizes: <D> = 13.5(1.1) nm and <D> = 62(12) nm. (c) Evolution of the nanoparticle-size distribution with the volume of trisodium citrate solution. (d) Comparing the maximum wavelength shifts from AuNPs with different sizes.

The influence of the chitosan upon the extinction efficiency of AuNPs was also studied for the set of spherical NP sizes. The relatively small spectral change for a given phantom confirmed the good dispersibility of the AuNPs. Systematically, the redshift was accompanied by the peak shifting from $\lambda_{max}$ = 505(1) nm in water, and $\lambda_{max}$ = 525(1) nm in chitosan. On the phantoms, the redshifts as a function of size were more pronounced when compared to colloidal AuNPs. The redshift dependences allowed the full optical determination of their properties, such as size, shape, and orientation. By increasing the value of refraction cores resulted in a longer wavelength redshift and narrower width of the absorption peak, which is in good agreement with the experiments [8, 18].

Through this means it was possible to derive an empirical formula to follow the absorption coefficients of AuNPs over the size range 10–100 nm in both environments, water versus chitosan, with refractive indexes of $n_1$ = 1.00 and $n_2$. We were able to use one equation to calculate the corresponding plasmon resonance peak position for AuNPs in chitosan phantoms compared to the reference water medium [18] [19] [20] [21]. Considering that the peak corresponding to the localized plasmon resonance depended on the refractive index of the surrounding medium, we proposed the following dependence:

$$\lambda_2 = \lambda_1 + (\lambda_1 - \lambda_0).(n_2-n_1) , \qquad (1)$$

since $\lambda_2$ represents the new position for the chitosan phantom media, with refractive index $n_2$, and $\lambda_1$ represents the wavelength position of the plasmon resonance peak for AuNPs in water, with refractive index $n_1$. The best coefficients after fitting Eq. (1) resulted in $n_2$ = 1.34 and $\lambda_0$ = 201 nm. It is worth noting that Eq. 1 is only valid for non-absorbing surrounding media, in which charge transfer should be considered in the plasmon bands shift. The absorption properties of the AuNPs embedded in chitosan medium were determined, which also showed the efficiency enhancement originating from plasmonic absorption coefficients.

As already commented in Fig. 2(c), the diameters changed with the volume of trisodium citrate solution, which causes the red shifting of the absorption peaks, as in Fig. 2(d). For the latter, the approximate linear behavior of data for the colloidal NPs allowed a reasonable prediction for the absorbance data following the Bouguer–Beer–Lambert -law [3]. To characterize scattering and absorption efficiency of a medium, reduced scattering ($\mu_s'$) and absorption ($\mu_a$) coefficients were introduced, which follow from the exponential law for light propagation in a layer of thickness d, I(d) = $I_0$ exp(-$\mu_{ext}$d), where I(d) is the intensity of transmitted light measured using the integrated sphere with a small aperture (on-line or collimated transmittance), $I_0$ is the incident light intensity, and $\mu_{ext}$ is the extinction or total attenuation coefficient, i.e., $\mu_{ext} = \mu_a + \mu_s'$. The absorption coefficient also follows the dependence: $\mu_a$ = k.C, where k is a constant and C the medium concentration. However, specifically in cloudy media with high scattering values such as the chitosan phantom, the extinction coefficient would not be adequate, since its curve $\lambda_{max}$ x <D> as in Fig. 2(d) deviated from linearity. Besides, in order to get appropriate values for the absorption coefficient, it was necessary to treat separately both the absorption and scattering, i.e., by observing their spectra dependence ($\mu_s'$ x $\lambda$) alone, more precisely the scattering mechanisms were obtained.

Thus, both the scattering and absorption spectra for the chitosan phantoms at different particle sizes were collected from the total diffuse reflectance and transmission. Figures 3(a) and 3(b) show the curves for <D> = 13.5(1.1) nm and <D> = 62(12), respectively. They were extracted by using the IAD method previously described, allowing the determination of the coefficients as a function of the incident wavelength. The coefficients were obtained by solving the problem for a model with infinite homogeneous medium in chitosan phantom, considering its refractive index $n_2$ = 1.34.

Figures 3(c) and 3(d) have showed that the reduced scattering coefficient ($\mu_s'$) of the phantoms were readily related to the AuNPs sizes. This also shows that the non-linearity of the coefficient $\mu_s'$ versus $\lambda_{max}$, because of the NP size effect, is one of the challenges in this work. By comparison, the linear dependence of this coefficient on the scatterers concentration with same sizes was observed in a previous study using polydimethylsiloxane phantoms [15]. As a mean of control, the characteristics of the scattering spectrum inside the chitosan phantom without embedded NPs were similar to each other, thereby meaning that only NP embedded phantoms were relevant to the scattering. To quantitatively estimate the wavelength-dependence of the scattering spectrum, the Mie theory was used to describe and relate the density of scattering objects within a volume A [16,22]. At the same time, the "slope" on the spectrum ($\mu_s' \times \lambda$) was related to the mean-size distribution of scattering, as an exponent $b_{Mie}$ in the relation below: $\mu_s'(Mie) = A \cdot (\lambda/\lambda_N)^{-b_{Mie}}$, where $\lambda_N$ represented the normalized wavelength, i.e., the corresponding wavelength divided by the peak maximum. Mie calculations were performed using a MATLAB code adapted for light scattering from concentric spheres on which plane-wave incidence was assumed to get the Mie coefficients. This assumption was justified by our use of the sub-diffraction limit sized AuNPs, and the fact that they were located far from the illuminating object if compared to the wavelength of visible light. Otherwise, Rayleigh scattering was dependent on $\lambda^{-4}$ and represented as $\mu_s'(Rayleigh) = B \cdot (\lambda/\lambda_N)^{-4}$ [23].

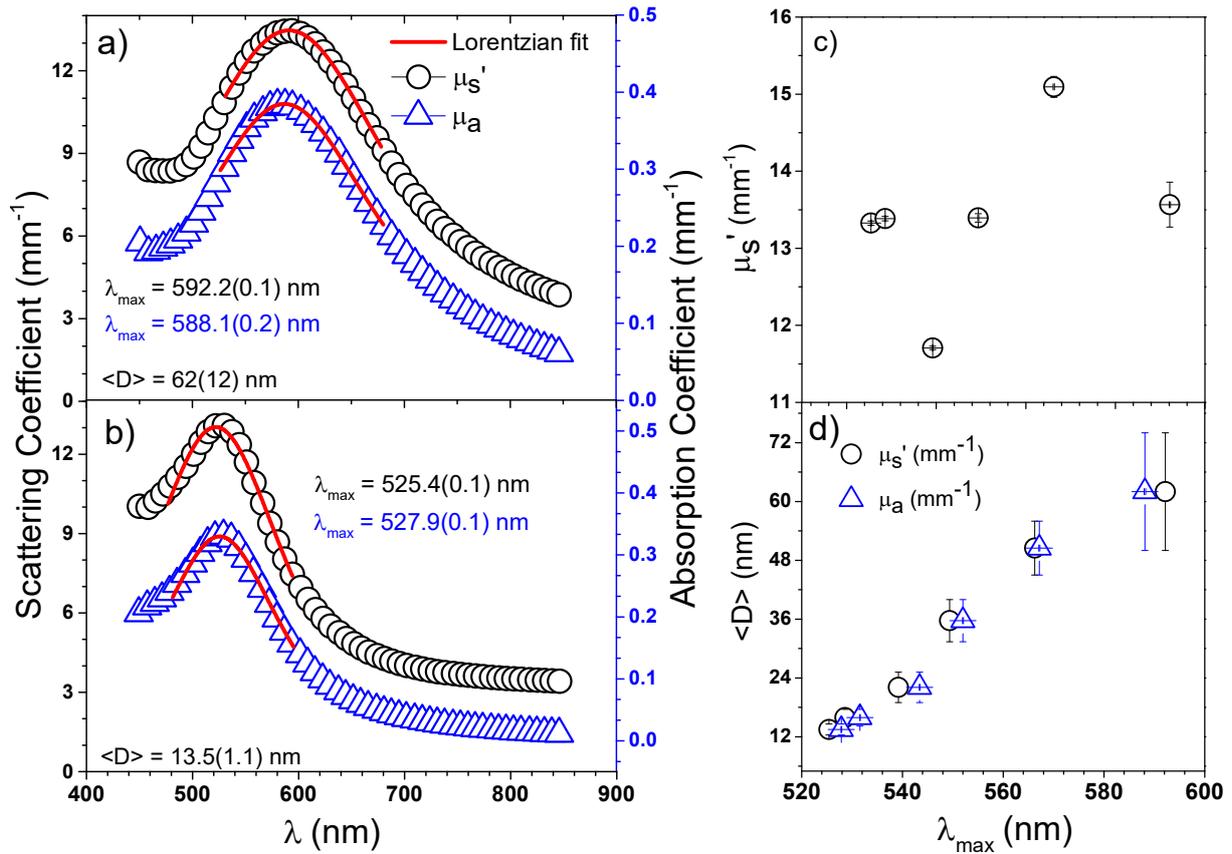

**Fig. 3** Absorption coefficient ($\mu_a$) and reduced scattering coefficient ($\mu_s'$) as a function of wavelength from AuNPs embedded in chitosan phantoms for different nanoparticle sizes: (a) <D> = 62(12) nm, (b) <D> = 13.5(1.1) nm. (c) Dependence of the reduced scattering coefficient with the maximum wavelength. (d) Evolution of the maximum wavelength of the absorption and reduced scattering as a function the AuNP diameter.

It is important to note that light scattering characteristics predicted by Mie theory approach Rayleigh scattering characteristics as the scattering particle becomes smaller. Then, the measured total reduced scattering coefficient spectra were then fitted through the use of a combination of Mie and Rayleigh scattering terms: $\mu_s'$(measured) = $\mu_s'$(Mie) + $\mu_s'$(Rayleigh). In other to access the correct scattering mechanism associated with the size of the AuNPs, we computed the Rayleigh scattering term as a fraction of the total measured reduced scattering coefficient, i.e., it was equal to $f_{Ray}$ multiplied by $\mu_s'$(Rayleigh). By doing this, the measured total reduced scattering coefficient was written as:

$$\mu_s'(\text{measured}) = C.[f_{Ray}(\lambda/\lambda_N)^{-4} + (1 - f_{Ray}).(\lambda/\lambda_N)^{-bMie}] . \quad (2)$$

The empirical curves of reduced scattering ($\mu_s'$ x $\lambda$, black circles), in Fig. 3, were fitted to Eq. 2, from the Mie Simulator algorithm based on the BHMIE code available from Bohren and Huffman [24]. This tool was capable of estimating the fitting parameters: $f_{Ray}$, bMie, and the constant C, being arranged in Table 1.

**Table 1** Exponential coefficients obtained from theory and associated scattering profile for different sizes of AuNPs relative to AuNPs embedded in chitosan phantoms.

| <D> (nm) | fRay | bMie | Dominant Scattering |
|---|---|---|---|
| 13.5(1.1) | 1 | 0.0 | Rayleigh |
| 15.9(1.7) | 1 | 0.0 | Rayleigh |
| 22.1(3.1) | 0.9 | 1.2 | Mie |
| 35.7(4.3) | 0.8 | 2.3 | Mie |
| 50.5(5.5) | 0.7 | 2.2 | Mie |
| 62(12) | 0.5 | 3.0 | Mie |

We were able to confirm our descriptions restricted to nearly spherical particles, with diameters varying between 13.5(1.1) and 62(12) nm. In the Mie theory, the scattering coefficient decays with the wavelength much more slowly as for $\lambda^{-b}$, with $1.2 < b < 3.0$, compared to $\lambda^{-4}$, from Rayleigh scattering. Therefore, this simple analysis determined the extinction, scattering, and absorption coefficients based on the Mie and Rayleigh scatterings. For AuNPs larger than 100 nm, outside this study, the formula did not contain a shape factor to account for non-spherical particles as well as clustering. The latter also causes perturbation of the plasmonic peak, as it was well known since the plasmon peak is sensitive to many parameters [2,3][19]. For robust and accurate measurements of AuNP concentrations, the discrete dipole approximation (DDA), as a numerical method, can be used to verify influences of particle sizes, interparticle distance, and even polarization direction of excited light on the optical properties of AuNPs embedded in phantom media [25].

**Conclusions**

In this work, the optical properties of AuNPs were readily tuned by varying its size and the surrounding medium. The plasmon resonances of the AuNPs were strongly sensitive to the NP sizes and the dielectric properties of the surrounding medium. To check this phenomenon, the absorption coefficients of a typical spectrum from a phantom containing only the AuNPs in water, and the spectrum from a phantom with AuNPs in chitosan, have been compared. In summary, we have quantified the changes in the absorption data anomalies, which includes broadening NP size-distribution after NPs were introduced into a phantom environment. Indeed, the redshift of the plasmon peak position in chitosan phantoms was larger when compared to AuNPs contained in water. The behavior of the effective optical scattering was observed for different sizes NPs, as the diameter size decreases the Rayleigh scattering becomes dominant over Mie scattering. It has been also demonstrated that the previous descriptions could fail for means other than water, for non-spherical particles, and aggregation for NPs greater than 100 nm in size. The main advantages of this study were biomedical oriented depending on circumstances, since the NP absorption and scattering spectra changing in a physiological environment may have a role on applications utilizing the optical properties of plasmonic NPs. For instance, in case of AuNPs being introduced into a physiological environment, they would interact with cells and their physicochemical properties might change as many proteins may adsorb on their surface, and the NPs may agglomerate.

## Acknowledgments

This work was supported by the Brazilian Agencies, CAPES, CNPq (308916/2015-8), and FAPEMIG (APQ-00582-14). The authors also thank to the Center of Advanced Microscopy in the acquisition of electron micrographies. RLS acknowledges the support from FAPEMIG-MG (APQ- 02378-17) and CAPES Foundation (Brazil) for grant EST-SENIOR-88881.119768/2016-01.

**Disclosures:** "The authors declare no conflicts of interest."